\documentstyle[editedvolume,numreferences,cite]{crckapb} 

\begin{opening}
\title{REARRANGEMENTS AND TUNNELING SPLITTINGS\protect\\
       IN SMALL WATER CLUSTERS}
\author{D.~J.~Wales}
\institute{University Chemical Laboratories\\
           Lensfield Road, Cambridge CB2 1EW, UK}
\end{opening}

\runningtitle{REARRANGEMENTS IN WATER CLUSTERS}

\begin{document}

% The \begin{document} command comes after the \end{opening}
% command.

\section{Introduction}

\def\water#1{$\rm (H_2O)_{#1}$}
\def\Dwater#1{$\rm (D_2O)_{#1}$}
\def\AB{{\it ab initio}}
\def\etal{{\it et al.\/}}
\def\hMS{h_{\rm MS}}
\def\hCNPI{h_{\rm CNPI}}
\def\hRM{h_{\rm RM}}

This paper was prepared in August 1997 for the proceedings volume of the NATO-ASI 
meeting on {\it Recent Theoretical and Experimental Advances in Hydrogen
Bonded Clusters\/} edited by Sotiris Xantheas, which has so far failed to
appear.
\bigskip

Recent far-infrared vibration-rotation tunneling (FIR-VRT) 
experiments \cite{FIR,PandS,SandB,Sreview} pose
new challenges to theory because the interpretation and prediction of such
spectra requires a detailed understanding of the potential energy surface (PES)
away from minima. In particular we need a global description of the PES
in terms of a complete reaction graph. Hence all the transition
states and associated mechanisms which might give rise to observable
tunneling splittings must be characterized. It may be possible to guess the
detailed permutations of atoms from the transition state alone, but experience
suggests this is unwise.

In this contribution a brief overview of the issues involved in treating 
the large amplitude motions of such systems will be given, with references
to more detailed discussions and some specific examples. In particular
we will consider the effective molecular symmetry group, the classification
of rearrangement mechanisms, the location of minima and transition states
and the calculation of reaction pathways. The application of these theories
to small water clusters ranging from \water{2}\ to \water{6}\ will then
be considered. More details can be found in recent reviews \cite{amvcd,Wiley,WWtri,TAMC}.

\section{The Effective Molecular Symmetry Group}

To classify the energy levels of systems executing large amplitude motions (non-rigid
or `floppy' molecules) we must look beyond point group symmetry. The PES 
contains all the permutational isomers of each stationary point. In the nomenclature
of Bone \etal\ \cite{Bone} we will refer to a {\it structure\/} as a particular molecular geometry and a {\it version\/} as a
particular labelled permutational isomer of a given structure.
Tunneling splittings occur when rovibronic wavefunctions localized in potential wells corresponding
to different permutational isomers interfere with each other. 

The Hamiltonian is invariant to arbitrary permutations of atoms of the same element, and to $E^*$,
the inversion of all coordinates through the space-fixed origin. Molecular energy levels may
therefore be classified according to irreducible representations (IR's) of the Complete Nuclear Permutation-Inversion
(CNPI) group which is the direct product of the inversion group and the group containing all possible permutations
of identical nuclei. The CNPI group is a true symmetry group of the
full molecular Hamiltonian in the absence of external fields, and the elements are usually
referred to as permutation-inversions (PI's). The CNPI group grows in size factorially
with the number of equivalent nuclei, and rapidly becomes unwieldy. 
Fortunately, Longuet-Higgins \cite{LH} showed that it is sufficient to consider the subgroup
of the CNPI group which contains only the PI's associated with barrierless processes and resolvable
tunneling splittings.
The associated rearrangement mechanisms are said to be {\it feasible\/}.
For rigid molecules the MS group, of order $\hRM$, is isomorphic to the point group \cite{Hougena,Hougenb},
and all the associated PI's correspond to barrierless processes. 
Each rovibronic level is $\hCNPI/\hRM$-fold degenerate because there are  $\hCNPI/\hRM$ distinct
versions on the PES \cite{Bone} and each one supports an identical stack of energy levels.

A feasible rearrangement with a non-zero barrier leads to an enlargement of the MS group
to include the new PI and all its products with the PI's of the rigid molecule MS group.
Further feasible mechanisms are included in a similar fashion.
The resulting MS group is a subgroup of the CNPI group obtained by removing all the PI's
for a given reference version which are not feasible. The order of the MS group, $\hMS$, 
and the order of the CNPI group, $\hCNPI$, must satisfy $h_{\rm CNPI}/h_{\rm MS}=M$, where
$M$ is an integer. The versions are then divided into $M$ disconnected sets. Each set
contains $\hMS/\hRM$ distinct versions \cite{Bone} all of which can be interconverted by one or more
feasible operations. 

To a first approximation the splitting pattern for any particular rovibronic level can be 
obtained from wavefunctions written as linear combinations of localized states which must
transform according to IR's of the MS group. In this approach we apply degenerate perturbation 
theory to obtain a secular determinant whose eigenfunctions and eigenvalues are the required
tunneling states and their energies. Mixing between different localized states is neglected
in this approximation. Since the localized wavefunctions decay
exponentially in the classically forbidden barrier regions it may be reasonable to make
a H\"uckel-type approximation including non-zero off-diagonal matrix elements, $\beta$, only for
minima which are directly connected by a feasible rearrangement. Even if the tunneling matrix
elements are unknown it is still possible to deduce the likely splitting pattern along
with symmetry classifications and nuclear spin statistics. This procedure has been
automated using a computer program to calculate the reaction graph from a minimal set of
generator PI's using double cosets \cite{HandR} and similarity transforms \cite{DJWpert}.

\section{Classification of Rearrangement Mechanisms}

We define a {\it degenerate\/} rearrangement mechanism as one which links permutational
isomers of the same structure via a single transition state \cite{degen}.
Here we follow Murrell and Laidler's definition of a transition state
as a stationary point with a single negative Hessian eigenvalue \cite{MandL}.
The largest tunneling splittings are expected for degenerate rearrangements (where the
localized states are in resonance) with low barriers, short paths and small effective
masses. Degenerate rearrangements can in turn be divided into two classes, namely
symmetric degenerate rearrangements (SDR's) where the two sides of the path are related
by a symmetry operation, and asymmetric degenerate rearrangements (ADR's) where they
are not \cite{Nourse}. Most of the rearrangements discussed in the following sections
are ADR's, which is likely to cause problems in choosing the fixed nodes for
excited state calculations in the diffusion Monte Carlo (DMC) approach \cite{DMC,GandCa,GandCb,GandCc}.

Recent experimental \cite{RJSiso} and theoretical \cite{Baciciso} studies of water trimer isotopomers
both indicate that significant tunneling splittings occur in these species, even when the
rearrangements involved are not strictly degenerate. (Within the Born-Oppenheimer approximation
the isotopomers have the same PES, but the vibronic states are not in resonance between different
wells.) We will also see in the following sections that tunneling splittings are observed for
\water{4}\ and \water{6}\ where no suitable degenerate rearrangements have been found
theoretically. In each case the tunneling may be the result of a series of steps mediated by
true transition states, or perhaps even higher index saddles, where the end points are true
permutational isomers. We suggest the term `indirect tunneling' to describe this situation
and distinguish it from `non-degenerate tunneling' where the interaction is between states
belonging to different structures.

% \begin{figure}
% \vspace{18.1cm}
% \caption{Geometries of the lowest energy TIP4P water clusters located
% by the `basin-hopping' algorithm.}
% \end{figure}

\section{Geometry Optimizations and Rearrangement Pathways}

Eigenvector-following provides a powerful technique for locating minima
and transition states and calculating reaction pathways \cite{CandM,SJTO,OTS,BASS,Bakera,Bakerb}.
The precise algorithms employed in the present work have been described in detail
elsewhere \cite{fiftyfive,WWpent}.
The step for eigendirection $i$ is
$$ h_i = { \pm2F_i\over |b_i|(1+\sqrt{1+4F_i^2/b_i^2}) }, \quad \left\{{\rm +\ for\ walking\ uphill\atop
\rm -\ for\ walking\ downhill}\right\}, $$
where $b_i$ is the corresponding Hessian eigenvalue and
$F_i$ is the component of the gradient in this direction.
A separate trust radius is employed for each eigendirection by comparing the
Hessian eigenvalue estimated from the current and previous gradients with the true value.
Analytic first and second derivatives of the energy were used at every step.

In the \AB\ calculations these derivatives were mostly generated by the CADPAC program \cite{CADPAC},
and Cartesian coordinates were used throughout along with projection to remove overall
rotation and translation \cite{BandH,PandM}.
Pathways were calculated by taking small displacements of $0.03\,a_0$ away from a transition state
both parallel and antiparallel to the transition vector, and then employing eigenvector-following
energy minimization to find the associated minimum. The pathways obtained by this procedure
have been compared to steepest-descent paths and pathways that incorporate a kinetic metric \cite{BandA}
in previous work---the mechanism is generally found to be correct \cite{WWtri}.
Calculations employing rigid body intermolecular potentials were performed using the ORIENT3 program \cite{PSW,WPS,WSP},
which contains the same optimization package adapted for center-of-mass/orientational coordinates.
Some of the calculations in the present work employ the 
ASP rigid water intermolecular potential of Millot and Stone \cite{MandS} (somewhat modified from the published version)
and the much simpler but widely-used TIP4P form \cite{Jorg,JCMIK}.

\begin{figure}
\vspace{13.0cm}
\caption{`Acceptor-tunneling' path calculated for \water{2}.
The first and last frames are the two minima, the middle frame is the transition state
and three additional frames on
each side of the path were selected to best illustrate the mechanism.
A suitably scaled transition vector is superimposed on the transition state; this
displacement vector lies parallel to the Hessian eigenvector corresponding
to the unique negative eigenvalue.}
\end{figure}

\section{Water Dimer}

\tolerance 5000
The pioneering experiment of Dyke, Mack and Muenter \cite{DMM} was followed
by many experimental and theoretical studies \cite{Fraser,Scheiner,PCLS} of \water{2}\ and \Dwater{2}.
Dyke \cite{Dyke} classified the rovibronic energy levels in terms of permutation-inversion
group theory, and Coudert and Hougen used their
internal-axis method and an empirical intermolecular potential to analyze the tunneling
splittings theoretically \cite{Hougen,CandHa,CandHb}.
A number of stationary points on the dimer PES were characterized by Smith \etal\ \cite{Smith},
including three true transition states. More recently the pathways corresponding to these
three feasible rearrangements have been characterized \cite{TAMC}.
Since the group theory of the water dimer is relatively well known, and has recently
been summarized with reference to the true pathways \cite{TAMC}, details will be
omitted here. Two points of interest arise for the `acceptor tunneling' pathway which
leads to the largest splittings. First, this pathway was found to correspond to a
`methylamine-type' process \cite{THIST} rather than a rotation about the local $C_2$
axis of the acceptor monomer (Figure 1) \cite{TAMC}.
This result is in agreement with the analysis of Pugliano \etal\ for the ground
state acceptor tunneling path based upon experiment \cite{Sdimer}.

\section{Water Trimer}

A flurry of experimental and theoretical studies \cite{Suzuki,liu,LandSb,bacic} followed the initial FIR-VRT
results of Pugliano and Saykally \cite{PnS}.
The energy level patterns observed in the most recent FIR-VRT experiments correspond to an
oblate symmetric rotor with 
a large negative inertial defect, implying extensive out-of-plane motion of the non-hydrogen-bonded hydrogens.
These results can be reconciled with the cyclic $C_1$ symmetry global minimum found in \AB\
calculations \cite{Bene,DJWtrimer} by vibrational averaging over large amplitude motions of the free (non-hydrogen-bonded)
hydrogens on the timescale of the FIR-VRT experiment.

The vibrational averaging which leads to the oblate symmetric top spectrum is caused
by the facile single {\it flip\/} mechanism. The corresponding 
transition state was probably first characterized by Owicki \etal\ \cite{Owicki} for an empirical potential.
We have illustrated the mechanism before \cite{amvcd,WWtri,DJWtrimer} and readers are referred to
these references for figures and a more detailed review of the literature.
The single flip mechanism links each permutational isomer to two others in cyclic sets of
six, giving a secular problem analogous to the $\pi$ system of benzene \cite{DJWtrimer,huckel}
with splitting pattern:
$$ 2\beta_{\rm f}(A_1), \qquad \beta_{\rm f}(E_2), \qquad -\beta_{\rm f}(E_1), \qquad -2\beta_{\rm f}(A_2),$$
where $\beta_{\rm f}$ is the tunneling matrix element for the flip.
The MS group has order six and is isomorphic to $C_{3h}$ \cite{PnS}.

The rearrangement responsible for the regular quartet splittings observed experimentally
is probably the analogue of the `donor tunneling' mechanism in the dimer, also known
as `bifurcation tunneling' because the transition state includes a bifurcated, double acceptor,
double-donor arrangement \cite{WWtri}.
When both the bifurcation mechanism and the single flip are feasible the MS group has order 48 \cite{DJWtrimer}.
However, there are six distinct ways for the bifurcation to occur with accompanying flips
of neighbouring monomers, and these give two distinct splitting patterns \cite{WWtri}.
The most recent experiments show that some of the quartets are further split by Coriolis coupling \cite{newtria,newtrib}.
Entirely regular quartets are found to be associated with generators containing $E^*$ in the
latter studies \cite{WWtri,newtrib}.

A more detailed account of the above mechanisms, group theory and quantum dynamics calculations
can be found elsewhere \cite{amvcd,WWtri}. The analysis of different bifurcation and flip
mechanisms is considered in more detail below for the water pentamer.

\section{Water Tetramer}

A symmetric doublet splitting of $5.6\,$MHz has been reported in two FIR-VRT experiments 
for \Dwater{4} \cite{teta,tetb}. The cyclic global minimum of the tetramer has $S_4$
symmetry \cite{Benetet,Scheragatet,Clementia,Clementib} and lacks the `frustration' 
exhibited by the cyclic trimer and pentamer.
Sch\"utz and coworkers found no true transition states correpsonding to degenerate
rearrangements in the torsional space of the tetramer \cite{Schutzteta,Schutztetb}.
A more systematic survey revealed many more non-degenerate rearrangement mechanisms
and just one true degenerate rearrangement of the global minimum, which disappeared
in correlated calculations \cite{WWtet}.

The effective MS group is isomorphic to $C_{2v}$ for an overall quadruple flip \cite{teta} 
mediated by any pathway which gives the same effective generator PI \cite{DandN}.
There are at least three possible routes involving (1) a concerted quadruple flip (via an
index four saddle), (2) true transition states and local minima, or (3) index two saddles
and local minima \cite{Schutzteta,WWtet}. The splitting pattern for the only direct
degenerate rearrangement found so far for this cluster is more complex \cite{WWtet}. 
Four-dimensional quantum calculations of torsional vibrational states
with a model Hamiltonian suggest that a stepwise pathway via the true transition states
may contribute most to tunneling \cite{WWtetb}, but there are too many approximations
involved for a definitive conclusion to be drawn. The possibility appears to remain that the
observed splittings are due to `indirect tunneling', as defined in \S3. A more detailed
discussion and illustrations of the mechanisms have been given elsewhere \cite{WWtet}.

\begin{figure}
\vspace{8cm}
\caption{The effect of A-bifurcation and A-bifurcation+B-flip on a reference version of
the water pentamer. The arrows indicate the direction of hydrogen-bond donation. The
generator PI is deduced by putting the structure in coincidence with the reference,
with an intervening inversion if necessary.}
\end{figure}

\begin{table}[htb]
\begin{center}
\caption{Generators and splitting patterns for the 15 distinct bifurcation+flip
combinations considered for water pentamer in \S 8. See also Figure 2.}
\begin{tabular}{lllc}
\hline
Description & generator & inverse & pattern \\
\hline
A-bif            & (ABCDE)(13579246810)* & E-bif           & A \\ % (AEDCB)(11086429753)* & A \\
A-bif+BCD-flips  & (AEDCB)(19753210864)  & B-bif+CDE-flips & A \\ % (ABCDE)(14681023579)  & A \\
E-bif+BCD-flips  & (ABCDE)(13571024689)  & D-bif+ABC-flips & A \\ % (AEDCB)(19864210753)  & A \\
D-bif+BCE-flips  & (ABCDE)(13581024679)  & C-bif+ABD-flips & A \\ % (AEDCB)(19764210853)  & A \\
B-bif+ACD-flips  & (AEDBC)(19754210863)  & C-bif+BCD-flips & A \\ % (ABCDE)(13681024579)  & A \\
A-bif+B-flip     & (ACEBD)(15937261048)  & D-bif+E-flip    & B \\ % (ADBEC)(18410627395   & B \\
A-bif+BC-flips   & (ADBEC)(17395284106)* & C-bif+DE-flips  & B \\ % (ACEBD)(16104825937)* & B \\
E-bif+D-flip     & (ADBEC)(17310628495)  & B-bif+A-flip    & B \\ % (ACEBD)(15948261037)  & B \\
E-bif+CD-flips   & (ACEBD)(17310628495)* & C-bif+AB-flips  & B \\ % (ADBEC)(17396284105)* & B \\
D-bif+EC-flips   & (ACEBD)(15938261047)* & B-bif+AC-flips  & B \\ % (ADBEC)(17410628395)* & B \\
A-bif+BCDE-flips & (12)*                 & self-inverse    & C \\ %                       & C \\
E-bif+ABCD-flips & (910)*                & self-inverse    & C \\ %                       & C \\
D-bif+ABCE-flips & (78)*                 & self-inverse    & C \\ %                       & C \\
B-bif+ACDE-flips & (34)*                 & self-inverse    & C \\ %                       & C \\
C-bif+ABDE-flips & (56)*                 & self-inverse    & C \\ %                       & C \\
\hline
\end{tabular}
\end{center}
\end{table}

\section{Water Pentamer}

The water pentamer exhibits a number of similarities to the trimer due to the frustrated
cyclic global minimum. The first FIR-VRT results for \Dwater{5}\ did not reveal any
tunneling splittings \cite{RJSpentamer}. However, analogues of the single flip and
bifurcation mechanisms, described above for the trimer, have been found along with a
number of pathways connecting higher energy minima \cite{WWpent,tales}.
Both the flip and the bifurcation mechanism considered in isolation produce an MS group
of order 10 isomorphic to $C_{5h}$ with each version connected to two others in a cyclic reaction graph containing
10 versions. The predicted splitting pattern in the simplest H\"uckel approximation is the same
as for the $\pi$-system of 10-annulene (cyclodecapentaene):
$$ 2\beta(A'), \quad \phi \beta(E_2''), \quad \phi^{-1}\beta(E_1'), \quad
    -\phi^{-1}\beta(E_1''), \quad -\phi \beta(E_2'), \quad -2\beta(A''),  $$
where $\beta$ is the appropriate tunneling matrix element,
$\phi=(\sqrt{5}+1)/2$ is the golden ratio and $\phi^{-1}=(\sqrt{5}-1)/2=1/\phi$.
The symmetry species in parentheses are appropriate if the generator corresponds
to the operation $S_5$ of $C_{5h}$. If both the flip and bifurcation mechanisms are feasible then
the MS group increases in dimension to order 320 and the splitting pattern becomes rather
complicated \cite{WWpent}.
Qualitative estimates for the magnitude of the splittings suggested that the flip should lead
to observable effects in both \water{5}\ and \Dwater{5}, while bifurcation tunneling might
produce splittings at the limit of current resolution \cite{WWpent}.
The latest experimental results appear to be in agreement with these predictions \cite{RJSpentamerb}.

\begin{table}[htb]
\begin{center}
\caption{Splitting pattern type C for five of the possible pentamer bifurcation/flip
combinations. $\phi=(\sqrt{5}+1)/2$.}
\begin{tabular}{clcclc}
\hline
level & \hfil energy & symmetry & level & \hfil energy & symmetry \\
\hline
   1   &    $2\beta_f+\beta_b$        & $A_1^+$ & 56 &$-2\beta_f-\beta_b$        & $A_1^-$ \\
   2   &    $2\beta_f+0.603\beta_b$   & $H_6^+$ & 55 &$-2\beta_f-0.603\beta_b$   & $H_6^-$ \\
   3   &    $2\beta_f+0.206\beta_b$   & $H_1^+$ & 54 &$-2\beta_f-0.206\beta_b$   & $H_1^-$ \\
   4   &    $2\beta_f+0.203\beta_b$   & $H_2^+$ & 53 &$-2\beta_f-0.203\beta_b$   & $H_2^-$ \\
   5   &    $2\beta_f-0.194\beta_b$   & $H_3^+$ & 52 &$-2\beta_f+0.194\beta_b$   & $H_3^-$ \\
   6   &    $2\beta_f-0.197\beta_b$   & $H_4^+$ & 51 &$-2\beta_f+0.197\beta_b$   & $H_4^-$ \\
   7   &    $2\beta_f-0.597\beta_b$   & $H_5^+$ & 50 &$-2\beta_f+0.597\beta_b$   & $H_5^-$ \\  
\medskip
   8   &    $2\beta_f-\beta_b$        & $A_2^+$ & 49 &$-2\beta_f+\beta_b$        & $A_2^-$ \\
   9   &    $\phi\beta_f+\beta_b$     & $E_4^-\oplus H_5^-$  & 48 &$-\phi\beta_f-\beta_b$     & $E_4^+\oplus H_5^+$ \\
  10   &    $\phi\beta_f+0.848\beta_b$& $H_3^-$  & 47 &$-\phi\beta_f-0.848\beta_b$& $H_3^+$ \\
  11   &    $\phi\beta_f+0.451\beta_b$& $H_1^-\oplus H_4^-$ & 46 &$-\phi\beta_f-0.451\beta_b$& $H_1^+\oplus H_4^+$ \\
  12   &    $\phi\beta_f+0.204\beta_b$& $H_5^-$  & 45 &$-\phi\beta_f-0.204\beta_b$& $H_5^+$ \\
  13   &    $\phi\beta_f+0.050\beta_b$& $H_2^-$  & 44 &$-\phi\beta_f-0.050\beta_b$& $H_2^+$ \\
  14   &    $\phi\beta_f-0.044\beta_b$& $H_4^-$  & 43 &$-\phi\beta_f+0.044\beta_b$& $H_4^+$ \\
  15   &    $\phi\beta_f-0.196\beta_b$& $H_6^-$  & 42 &$-\phi\beta_f+0.196\beta_b$& $H_6^+$ \\
  16   &    $\phi\beta_f-0.444\beta_b$& $H_2^-\oplus H_3^-$ & 41 &$-\phi\beta_f+0.444\beta_b$& $H_2^+\oplus H_3^+$ \\
  17   &    $\phi\beta_f-0.846\beta_b$& $H_1^-$  & 40 &$-\phi\beta_f+0.846\beta_b$& $H_1^+$ \\
\medskip
  18   &    $\phi\beta_f-\beta_b$     & $E_2^-\oplus H_6^-$  & 39 &$-\phi\beta_f+\beta_b$     & $E_2^+\oplus H_6^+$ \\
  19   &     $\phi^{-1}\beta_f+\beta_b$     & $E_1^+\oplus H_6^+$  & 38 &$-\phi^{-1}\beta_f-\beta_b$     & $E_1^-\oplus H_6^-$  \\
  20   &     $\phi^{-1}\beta_f+0.847\beta_b$& $H_2^+$  & 37 &$-\phi^{-1}\beta_f-0.847\beta_b$& $H_2^-$  \\
  21   &     $\phi^{-1}\beta_f+0.451\beta_b$& $H_1^+\oplus H_4^+$ & 36 &$-\phi^{-1}\beta_f-0.451\beta_b$& $H_1^-\oplus H_4^-$  \\
  22   &     $\phi^{-1}\beta_f+0.201\beta_b$& $H_6^+$  & 35 &$-\phi^{-1}\beta_f-0.201\beta_b$& $H_6^-$  \\
  23   &     $\phi^{-1}\beta_f-0.042\beta_b$& $H_3^+$  & 34 &$-\phi^{-1}\beta_f+0.042\beta_b$& $H_3^-$  \\
  24   &     $\phi^{-1}\beta_f-0.053\beta_b$& $H_1^+$  & 33 &$-\phi^{-1}\beta_f+0.053\beta_b$& $H_1^-$  \\
  25   &     $\phi^{-1}\beta_f-0.199\beta_b$& $H_4^+$  & 32 &$-\phi^{-1}\beta_f+0.199\beta_b$& $H_5^-$  \\
  26   &     $\phi^{-1}\beta_f-0.444\beta_b$& $H_2^+\oplus H_3^+$ & 31 &$-\phi^{-1}\beta_f+0.444\beta_b$& $H_2^-\oplus H_3^-$  \\
  27   &     $\phi^{-1}\beta_f-0.847\beta_b$& $H_4^+$  & 30 &$-\phi^{-1}\beta_f+0.847\beta_b$& $H_4^-$  \\
  28   &     $\phi^{-1}\beta_f-\beta_b$     & $E_3^+\oplus H_5^+$  & 29 &$-\phi^{-1}\beta_f+\beta_b$     & $E_3^-\oplus H_5^-$  \\
\hline
\end{tabular}
\end{center}
\end{table}

Since new experiments suggest that bifurcation tunneling splittings might be resolvable 
for \water{5}\ we will now consider the group theory in more detail to illustrate how
such analyses are performed. The labelling scheme and assignment of generator PI's
is illustrated for two possible bifurcation+flip combinations in Figure 2. The
labelling scheme in the figure enables us to identify all the possible bifurcation/flip
processes which correspond to degenerate rearrangements using the letters A--E to
label the five monomers. Here we exclude processes where the bifurcating
monomer also flips. The bifurcation can occur at any of the five monomers, and in each case
the frustrated pair of neighbouring free hydrogens can appear on any of the five edges.
This gives a total of 25 bifurcation/flip combinations. Five of the resulting generators
are self-inverse while the other twenty fall into non-self-inverse pairs to give a
total of 15 distinct mechanisms (Table 1). For the same labelling scheme the generator
for the single flip is (ABCDE)(13579)(246810)* with inverse (AEDCB)(19753)(210864)*.

Previously we identified two splitting patterns for different bifurcation/flip combinations
in the pentamer, and labelled these A and B. In the present systematic analysis a third
splitting pattern, type C, emerges, as shown in Table 2. All 15 combinations result in
the same MS group, $G(320)$, described previously \cite{WWpent}. The C-type pattern occurs for the
five generators of the form $(ij)^*$; for the trimer such generators also correspond to
one particular splitting pattern \cite{WWtri}. The A-type pattern is found for the bifurcation
accompanied by zero or three flips and the B-type pattern is found for the bifurcation
accompanied by one or two flips (Table 1). All three patterns include four sets of ten lines and
two sets of eight; they all involve irregular submultiplets.
In the C pattern every energy level $\lambda$ has a partner with energy $-\lambda$ (and opposite parity with
respect to $E^*$), implying that
there are no odd-membered rings in the reaction graph \cite{CandR}.
The splittings within each manifold with a constant coefficient of $\beta_f$ (the tunneling
matrix element for the flip) also obey an approximate pairing rule.

\begin{figure}
\vspace{11.5cm}
\caption{Single flip mechanism which interconverts hexamer cage isomers C1 and C2.}
\end{figure}

\section{Water Hexamer}

On the basis of an isotope mixture test Liu \etal\ \cite{hexamer} have
assigned a VRT band of \water{6}\ at $83\,$cm$^{-1}$. For this cluster
a number of different structures lie rather close in energy, with the
four lowest energy isomers reported by Tsai and Jordan separated by only 
about $100\,$cm$^{-1}$ \cite{TandJ}.
The most accurate \AB\ calculations performed to date suggest that a `cage' structure lies lowest, followed
closely by `prism' and `book' forms \cite{KJZ}.
The assignment of the experimental spectrum to the cage isomer was made on 
the basis of DMC calculations of the vibrationally averaged rotational constants \cite{hexamer}.
It has been suggested \cite{hexamer} that a single structure is observed because the clusters have
very low internal temperatures (around 5$\,$K) under experimental conditions.
However, each experimental line is split into a triplet with intensity ratio 9:6:1 and
separation $1.92\,$MHz.
Liu \etal\ have explained how this pattern might arise from two hypothetical degenerate
rearrangements of monomers in similar chemical environments \cite{hexamer}.
However, no low-lying degenerate rearrangements have been found in studies employing
both empirical and \AB\ calculations \cite{TAMC,GandCamv,GandChex}.
Four isomers of the cage have been found independently with very similar ASP-type
potentials \cite{TAMC,GandCamv,GandChex} and checked in \AB\ calculations \cite{GandCamv,GandChex}.
Flip and bifurcation rearrangements of the terminal single-donor, single-acceptor monomers
exist which interconvert 16 versions of the cage structure,
including four versions of each low-lying isomer \cite{TAMC}. 
Two examples are shown in Figures 3 and 4.

\begin{figure}
\vspace{11.5cm}
\caption{Bifurcation mechanism which interconverts hexamer cage isomers C1 and C2.}
\end{figure}

The resulting MS group can be found by considering the effective
generators for combinations of flip and bifurcation processes which together result in
a permutation of the same structure \cite{DandN}.
The same group is obtained for all four of the cage isomers;
it contains four elements and is isomorphic to $C_{2v}$ \cite{TAMC}.
The effective reaction graph for one of the four cage isomers is shown in Figure 5.
The four versions are connected in a cyclic reaction
graph so that the simplest H\"uckel $\pi$ treatment gives a splitting pattern equivalent to
that of the $\pi$ system in cyclobutadiene:
$$ 2\beta (A_1), \qquad 0 (A_2,B_1), \qquad -2\beta (B_2), $$
where the symmetry labels are appropriate for a particular correspondence between the PI's and the
elements of $C_{2v}$ \cite{TAMC}.
The accidental degeneracy of the $A_2$ and $B_1$ states would be broken at higher resolution because
the $\beta$ matrix elements connecting the four versions of each isomer are all slightly different.
The relative nuclear spin weights for rovibronic states are 9:3:3:1 for \water{6}\ and
4:2:2:1 for \Dwater{6} corresponding to
$A_1$:$A_2$:$B_1$:$B_2$. If the accidental degeneracy is unresolved then the relative intensities
of the three triplet components would be 9:6:1 for \water{6}\ and 4:4:1 for \Dwater{6}.
This result is equivalent to that obtained by Liu \etal\ \cite{hexamer} but does not
require the existence of any hypothetical mechanisms. A similar analysis holds for
all four cage isomers. However, if this explanation is correct then it seems likely that all four cage isomers
would be present under experimental conditions, which seems to be incompatible with the interpretation
of the spectrum in terms of a single isomer \cite{hexamer}. Unless low energy degenerate
rearrangements of the cage isomers are found the above `indirect tunneling' mechanism
seems to be the most plausible explanation that theory can provide.

\begin{figure}
\vspace{4.25cm}
\caption{The effective reaction graph for four permutational isomers of cage
structure C1 which are linked by stepwise flip and bifurcation rearrangements
involving different cage structures.}
\end{figure}

Two-dimensional quantum calculations of torsional vibrational states
for the single-acceptor, single-donor monomers have also been carried out \cite{TAMC}.
The other four monomers are held fixed in this model. Both TIP4P \cite{Jorg,JCMIK} and
ASP potentials \cite{MandS} were considered. For TIP4P all the wavefunctions were
delocalized over the torsional degrees of freedom of the two monomers. However, for
the ASP potential the ground state was found to be localized, in agreement with
DMC calculations of Gregory and Clary \cite{GandChex}. The latter result appears to
be incompatible with the experimental observation of tunneling splittings. However,
the two-dimensional calculation suggests that there exist low-lying vibrational
states which are delocalized over more than one cage isomer. More details, along
with illustrations of the wavefunctions, can be found elsewhere \cite{TAMC}.
These results provide a clear illustration of how tunneling splittings can be
sensitive to the excitation of particular vibrational modes. Since experiment measures
the difference between ground and excited state splittings a direct comparison
with theory is often difficult.

% acknowledge Matt and RS

\def\jcp{{\it J.~Chem.~Phys.}}
\def\pra{{\it Phys.~Rev.~A}}
\def\jpc{{\it J.~Phys.~Chem.}}
\def\jpca{{\it J.~Phys.~Chem.~A}}
\def\molphys{{\it Molec.~Phys.}}
\def\cp{{\it Chem.~Phys.}}
\def\cpl{{\it Chem.~Phys.~Lett.}}
\def\ss{{\it Surface Sci.}}
\def\prb{{\it Phys.~Rev.~B}}
\def\pre{{\it Phys.~Rev.~E}}
\def\acp{{\it Adv.~Chem.~Phys.}}
\def\prl{{\it Phys.~Rev.~Lett.}}
\def\fdcs{{\it Faraday Discuss.~Chem.~Soc.}}
\def\jms{{\it J.~Mol.~Spec.}}
\def\ica{{\it Inorg.~Chim.~Acta}}
\def\tca{{\it Theo.~Chim.~Acta}}
\def\faradayII{{\it J.~Chem.~Soc., Faraday II}}
\def\fartrans{{\it J.~Chem.~Soc., Faraday Trans.}}
\def\tetlet{{\it Tetrahedron Letters}}
\def\jacs{{\it J.~Amer.~Chem.~Soc.}}
\def\ijqc{{\it Int.~J.~Quantum Chem.}}
\def\poly{{\it Polyhedron}}
\def\joc{{J.~Organomet.~Chem.}}
\def\dalton{{\it J. Chem. Soc., Dalton Trans.}}
\def\pnas{{\it Proc.~Natl.~Acad.~Sci.~USA}}
\def\chemcomm{{\it J.~Chem.~Soc., Chem.~Comm.}}
\def\ic{{\it Inorg.~Chem.}}
\def\sci{{\it Science}}

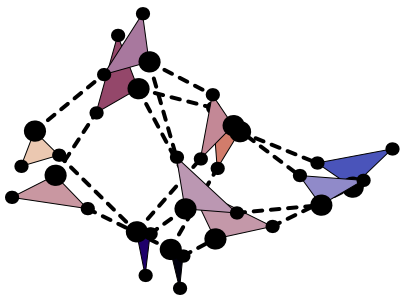
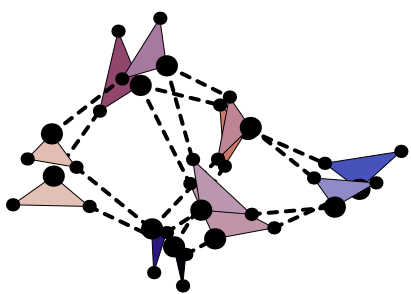
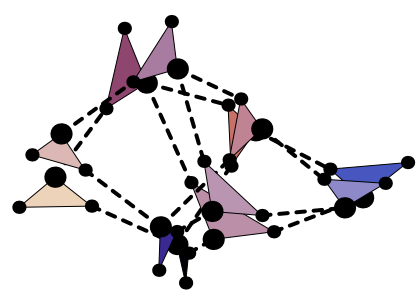
\end{document}